\documentclass[a4paper,11pt]{article}
\usepackage{epsfig}
\usepackage{amssymb}
\usepackage{amsmath}

\title{Non trivial spacetime effects in a supersymmetric model}
\author{V.K. Oikonomou\thanks{voiko@physics.auth.gr}\\
Department of Theoretical Physics, Aristotle University of Thessaloniki,\\
Thessaloniki 541 24 Greece}
\begin{document}
\maketitle

\begin{abstract}
We study a ${N=1}$ supersymmetric model in a ${S}^{1}{\times R}^{3}$
spacetime. We find  that by choosing appropriate boundary conditions
for the contributing fields supersymmetry can be preserved. However
if we add a hard supersymmetry breaking term, we observe that for
small values of the length of the ${S}^{1}$ dimension, supersymmetry
remains unbroken and breaks spontaneously when the length exceeds a
critical value. The final picture resembles the first order phase
transition picture.
\end{abstract}

\section*{\protect \Large Introduction}

In general,\ when studying supersymmetric theories in flat
spacetime, the background metric is assumed to be ordinary
Minkowski. Spacetime topology may affect the boundary conditions of
the fields that are integrated in the path integral.\ Given a class
of metrics, several spacetime topologies are allowed.\ Here we shall
focus on a model that has ${S}^{1}{\times R}^{3}$ topology
underlying the spacetime, $S^{1}$ refers to a spatial dimension.\
The specific topology is a homogeneous topology of the flat
Clifford--Klein type \cite{ellis}.\ Non trivial topology, implies
non trivial field configurations, that enter dynamically in the
action.\ We shall investigate their impact using the effective
potential method \cite{dolan,coleman weinberg}.\ It is known that
when supersymmetric theories are studied in non trivial spacetimes
(or at finite temperature), supersymmetry is, in general,
spontaneously broken.\ This is due to the appearance of vacuum terms
which have different coefficients for fermions and bosons.\ As a
result, the effective potential of the theory has no longer its
minimum at zero thus, supersymmetry is broken.\ This quite general
phenomenon can only be avoided, if in some way these vacuum terms
are canceled \cite{love}.

The existence of non trivial field configurations due to non trivial
topology (twisted fields), was first pointed out by Isham
\cite{isham} and then adopted by others
\cite{gongcharov,ford,spalluci}.\ In our case, the topological
properties of $S^{1}{\times R}^{3}$ are classified by the first
Stieffel class $H^{1}(S^{1}{\times R}^{3},Z_{\widetilde{2}})$ which
is isomorphic to the singular (simplicial) cohomology group
${H}_{1}({S} ^{1}{\times R}^{3}{,Z}_{2})$ because of the triviality
of the ${Z}_{\widetilde{2}}$ sheaf.\ Now, it is known that
$H^{1}{(S}^{1}{\times R}^{3}{,Z}_{\widetilde{2}}{)=Z}_{2}$
classifies the twisting of a bundle.\ To be exact, it describes and
classifies the orientability of a bundle globally.\ In our case, the
classification group is ${Z}_{2}$ and, we have two locally
equivalent bundles, which are however different globally
{\it{i.e.}}~cylinder-like and moebius strip-like.\ The mathematical
lying behind, is to find the sections that correspond to these two
bundles, classified completely by $Z_{2}$ \cite{isham}.\ The
sections we shall consider are real scalar fields and Majorana
spinor fields.\ These fields carry a topological number called
moebiosity (twist), which distinguishes between twisted and
untwisted fields.\ The twisted fields obey anti-periodic boundary
conditions, while untwisted fields periodic in the compactified
dimension (see below).\ Normally one takes scalars to obey periodic
and fermions anti-periodic boundary conditions, disregarding all
other configurations that may arise from non trivial topology.\ Here
we shall consider all these configurations.\ If $\varphi _{u}$,
$\varphi_{t}$ and $\psi _{t}$, $\psi _{u}$ denote the untwisted and
twisted scalar and twisted and untwisted spinor respectively, then
the boundary conditions in the ${S}^{1}$ dimension are, $\varphi
_{u}(x,0)=\varphi _{u}(x,L)$ and $\varphi _{t}(x,0)=-\varphi _{t}(x,L)%
$ for scalars and $\psi _{u}(x,0)=\psi _{u}(x,L)$, $\psi%
_{t}(x,0)=-\psi _{t}(x,L)$ for fermions, where $x$ stands for the
remaining two spatial  and one time dimensions  not affected by the
boundary conditions.\ Spinors (both Dirac and Majorana), still
remain Grassmann quantities.\ We assign the untwisted fields twist
${h}_{0}$  (the trivial element of ${Z}_{2})$ and the twisted fields
twist $h_{1}$  (the non
trivial element of ${Z}_{2}$).\ Recall that $h_{0}+h_{0}=h_{0}$ ($0+0=0$), $%
h_{1}+h_{1}=h_{0}$ ($1+1=0$), $h_{1}+h_{0}=h_{1}$ ($1+0=1$).\ We
require the Lagrangian to have ${h}_{0}$ moebiosity.\ The
topological charges flowing at the
interaction vertices must sum to ${h}_{0}$ under ${H}^{1}{(S}^{1}{\times R}^{3}{,Z%
}_{\widetilde{2}}{)}$.\ For supersymmetric models, supersymmetry
transformations impose some restrictions on the twist assignments of
the superfield component fields \cite{gongcharov}.

Grassmann fields cannot acquire vacuum expectation value (vev) since
we require the vacuum value to be a scalar representation of the
Lorentz group.\ Thus, the question is focused on the two scalars.\
The twisted scalar cannot acquire non\ zero\ vev\ \cite{ford},
consequently, only untwisted scalars are allowed to develop vev's.

The model under consideration, is described by the global ${N=1}$,
${d=4}$ supersymmetric Lagrangian,%
\begin{equation}
\mathcal{L}=[\Phi _{1}^{+}\Phi _{1}]_{D}+[\Phi ^{+}\Phi
]_{D}+[\frac{m_{1}}{2}\Phi ^{2}+\frac{g_{1}}{6}\Phi
^{3}+\frac{m}{2}\Phi _{1}^{2}+g\Phi \Phi _{1}^{2}]_{F}+{\rm{H.c}},
\label{lagra}
\end{equation}%
where $\Phi _{1}$, $\Phi $ are chiral superfields. In the above,%
\begin{eqnarray}
\Phi &=&\varphi _{u}(x)+\sqrt{2}\theta \psi _{u}(x)+\theta \theta
F_{\varphi _{u}}+i\partial _{\mu }\varphi _{u}(x)\theta \sigma ^{\mu }\bar{\theta } \\
&&-\frac{i}{\sqrt{2}}\theta \theta \partial _{\mu }\psi _{u}(x)\sigma ^{\mu }%
\bar{\theta }-\frac{1}{4}\partial _{\mu }\partial ^{\mu }\varphi
_{u}^{+}(x)\theta \theta \bar{\theta }\bar{\theta }, \notag
\end{eqnarray}%
is a left chiral superfield.\ It contains the untwisted scalar field
components and the untwisted Weyl fermion. Although the untwisted
scalar is complex, we shall use the real components which will be
the representatives of the
sections of the trivial bundle classified by ${H}^{1}{(S}^{1}{\times R}^{3}{%
,Z}_{\widetilde{2}}{)}$. Moreover,%
\begin{eqnarray}
\Phi _{1} &=&\varphi _{t}(x)+\sqrt{2}\theta \psi _{t}(x)+\theta
\theta F_{\varphi _{t}}+i\partial _{\mu }\varphi _{t}(x)\theta
\sigma ^{\mu }\bar{\theta }
\\
&&-\frac{i}{\sqrt{2}}\theta \theta \partial _{\mu }\psi _{t}(x)\sigma ^{\mu }%
\bar{\theta }-\frac{1}{4}\partial _{\mu }\partial ^{\mu }\varphi
_{t}^{+}(x)\theta \theta \bar{\theta }\bar{\theta }, \notag
\end{eqnarray}%
is another left chiral superfield containing the twisted scalar field and
the twisted Weyl fermion. Writing down (\ref{lagra}) in component form, we
get (writing Weyl fermions in the Majorana representation):%
\begin{align}
\mathcal{L}& =\partial _{\mu }\varphi _{u}^{+}\partial ^{\mu
}\varphi _{u}-\left \vert m_{1}\varphi _{u}+\frac{g_{1}}{2}\varphi
_{u}\varphi _{u}+g\varphi _{t}^{2}\right \vert ^{2}+i\overline{\Psi
}_{t}\gamma ^{\mu }\partial _{\mu
}\Psi _{t}-\frac{1}{2}m\overline{\Psi }_{t}\Psi _{t} \label{test}\\
& -\frac{g_{1}}{4}(\overline{\Psi }_{u}\Psi _{u}-\overline{\Psi
}_{u}\gamma
_{5}\Psi _{u})\varphi _{u}-\frac{g_{1}}{4}(\overline{\Psi }_{u}\Psi _{u}+%
\overline{\Psi }_{u}\gamma _{5}\Psi _{u})\varphi _{u}^{+}+\partial _{\mu
}\varphi _{t}^{+}\partial ^{\mu }\varphi _{t}-  \notag \\
& \left \vert m\varphi _{t}+2g\varphi _{t}\varphi _{u}\right \vert ^{2}+i%
\overline{\Psi }_{u}\gamma ^{\mu }\partial _{\mu }\Psi _{u}-\frac{1}{2}m_{1}%
\overline{\Psi }_{u}\Psi _{u}-  \notag \\
& \frac{g}{4}(\overline{\Psi }_{t}\Psi _{t}-\overline{\Psi }_{t}\gamma
_{5}\Psi _{t})\varphi _{u}-\frac{g}{4}(\overline{\Psi }_{t}\Psi _{t}+%
\overline{\Psi }_{t}\gamma _{5}\Psi _{t})\varphi _{u}^{+}.  \notag
\end{align}%
We can explicitly check that moebiosity is conserved at all
interaction vertices {\it{i.e.}}~equals ${h}_{0}$.\ The moebiosity
of\ $\varphi _{u}$ and $\Psi _{u}$ is ${h}_{0}$ while for $\varphi
_{t}$\ and\ $\Psi _{t}$ is ${h}_{1}$.\ One can use the ${Z}_{2}$
cyclic group properties to prove that the Lagrangian (\ref{test})
has moebiosity ${h}_{0}$.\ The complex field $\varphi _{u}$ can be
written in terms of real components as ${\varphi }_{u}{=\chi
+i\varphi }_{u_{2}}{/}\sqrt{2}$, where $\chi ={v+(\varphi
}_{u_{1}}{)/}\sqrt{2}$ (${v}$ is the classical value).\ Thus,
$\varphi _{u_{1}}$\ and $\varphi _{u_{2}}$ are
real untwisted field configurations belonging to the trivial element of ${%
H}^{1}{(S}^{1}{\times R}^{3}{,{Z}_{\widetilde{2}})}$ and satisfying
periodic boundary conditions in the compactified dimension.\ The
twisted scalar field
can be written as ${\varphi }_{t}{=(\varphi }_{t_{1}}{+i\varphi }_{t_{2}}{)/}%
\sqrt{2}$, since, this field, being a member of the non trivial
element of ${H}^{1}{(S}^{1}{\times R}^{3}{,Z}_{\widetilde{2}}{)}$
cannot acquire a vev. The tree order masses of the two Majorana
fermion
fields and the four bosonic fields are calculated to be:%
\begin{eqnarray}
m_{b_{1}}^{2} &=&m_{1}^{2}+3g_{1}m_{1}v+3g_{1}^{2}v^{2}/2 \label{mass}\\
\ m_{b_{2}}^{2} &=&m_{1}^{2}+g_{1}m_{1}v+g_{1}^{2}v^{2}/2  \notag \\
m_{t_{1}}^{2}\bigskip &=&m^{2}+4gmv+4g^{2}v^{2}+g^{2}m_{1}v/\sqrt{2}%
+g^{2}g_{1}v^{2}/4\   \notag \\
\ m_{t_{2}}^{2} &=&m^{2}+4gmv-g^{2}m_{1}v/\sqrt{2}-g^{2}g_{1}v^{2}/4  \notag
\\
m_{f_{1}} &=&m_{1}+g_{1}v,\ m_{f_{2}}=m+2gv.  \notag
\end{eqnarray}%
In (\ref{mass}) $m_{b_{1}}$, $m_{b_{2}}$ are the masses of the untwisted bosons, $%
m_{t_{1}}$, $m_{t_{2}}$ are the masses of the twisted bosons and, finally, $m_{f_{1}}$%
, $m_{f_{2}}$ are the untwisted Majorana fermion and twisted
Majorana fermion masses respectively.\ We can check that the general
tree level result for theories with rigid supersymmetry in terms of
chiral superfields is satisfied (see \cite{martin}) {\it{i.e.}}~:
\begin{equation}
STr(M^{2})=\sum \limits_{j}(-1)^{2j}(2j+1)m_{j}^{2}=0.  \label{str}
\end{equation}%
Also, the following relations hold true:%
\begin{equation}
m_{b_{1}}^{2}+m_{b_{2}}^{2}=2m_{f_{1}}^{2},\
m_{t_{1}}^{2}+m_{t_{2}}^{2}=2m_{f_{2}}^{2}.\label{sp}
\end{equation}%
Since twisted scalars cannot acquire vacuum expectation value,
supersymmetry is not spontaneously broken at tree level, like in the
O' Raifeartaigh models. The auxiliary field equations,%
\begin{eqnarray}
F_{\varphi _{u}}^{+} &=&m_{1}\varphi _{u}+\frac{g_{1}}{2}\varphi
_{u}^{2}+g\varphi _{t}^{2}=0 \\
F_{\varphi _{t}}^{+} &=&m\varphi _{t}+2g\varphi _{u}\varphi _{t}=0,
\notag
\end{eqnarray}%
imply that $\varphi _{u}=0$ and $\varphi _{t}=0$, thus, at tree
level, no spontaneous supersymmetry breaking occurs.

We now proceed by assuming that the topology is changed to
${S}^{1}{\times R}^{3}$, while the local geometry remains Minkowski.
The metric is:
\begin{equation}
\mathrm{d}s^{2}=\mathrm{d}t^{2}-\mathrm{d}x_{1}^{2}-\mathrm{d}x_{2}^{2}-\mathrm{d}x_{3}^{2},
\end{equation}%
with $-\infty <x_{2},x_{3},t<\infty $ and $0<x_{1}<L$ with the points${\ x}%
_{1}=0~$and ${x}_{1}=L$ periodically identified.\ The boundary
conditions for the fields are:%
\begin{align}
~\varphi _{u}(x_{1},x_{2},x_{3},t) &=~~~~~\varphi _{u}(x_{1}+L,x_{2},x_{3},t) \\
~\varphi _{t}(x_{1},x_{2},x_{3},t) &=~{}-\varphi
_{t}(x_{1}+L,x_{2},x_{3},t)
\notag \\
~\Psi _{u}(x_{1},x_{2},x_{3},t) &=~~~~~\Psi
_{u}(x_{1}+L,x_{2},x_{3},t) \notag
\\
~\Psi _{t}(x_{1},x_{2},x_{3},t) &=~~-\Psi
_{t}(x_{1}+L,x_{2},x_{3},t). \notag
\end{align}%
In order to calculate the effective potential of the theory, we Wick
rotate the time direction ${t}\rightarrow {it}$\ thus giving the
background metric the Euclidean signature \cite{spalluci}.\ The
twisted fermions and twisted bosons will be summed over odd
Matsubara frequencies, while the untwisted fermions and untwisted
scalars will be summed over even Matsubara frequencies
\cite{dolan,coleman weinberg}.\ We shall adopt the
$\overline{DR}^{\prime }$ renormalization scheme \cite{martin}. The
Euclidean effective potential
at one loop level is:%
\begin{align}
V& =V_{0}+\frac{1}{64\pi ^{2}L}\sum \limits_{n=-\infty }^{\infty }\int \frac{%
\mathrm{d}^{3}k}{(2\pi )^{3}}{\bigg (}\ln [k^{2}+\frac{4\pi^{2} n^{2}}{L^{2}}%
+m_{b_{1}}^{2}] \\
& -2\ln [k^{2}+\frac{4\pi^{2} n^{2}}{L^{2}}+m_{f_{1}}^{2}]+\ln [k^{2}+\frac{%
4\pi^{2} n^{2}}{L^{2}}+m_{b_{2}}^{2}]  \notag \\
& -2\ln [k^{2}+\frac{\pi^{2} (2n+1)^2}{L^{2}}+m_{f_{2}}^{2}]+\ln [k^{2}+\frac{%
\pi^{2} (2n+1)^2}{L^{2}}+m_{t_{1}}^{2}]  \notag \\
& +\ln [k^{2}+\frac{\pi^{2} (2n+1)^2}{L^{2}}+m_{t_{2}}^{2}]{\bigg
).} \notag
\end{align}%
${V}_{0}~$includes the tree and the one loop continuum corrections,%
\begin{align}
V_{0}& =m_{1}^{2}v^{2}+g_{1}^{2}m_{1}v^{3}+\frac{g_{1}^{2}v^{4}}{4}+\frac{1}{%
64\pi ^{2}}{\bigg (}m_{b_{1}}^{4}(\ln [\frac{m_{b_{1}}^{2}}{\mu ^{2}}]-%
\frac{3}{2}) \\
& +m_{b_{2}}^{4}(\ln [\frac{m_{b_{2}}^{2}}{\mu ^{2}}]-\frac{3}{2}%
)+m_{t_{1}}^{4}(\ln [\frac{m_{t_{1}}^{2}}{\mu ^{2}}]-\frac{3}{2}%
)+m_{t_{2}}^{4}(\ln [\frac{m_{t_{2}}^{2}}{\mu ^{2}}]-\frac{3}{2})  \notag \\
& -2m_{f_{1}}^{4}(\ln [\frac{m_{f_{1}}^{2}}{\mu ^{2}}]-\frac{3}{2}%
)-2m_{f_{2}}^{4}(\ln [\frac{m_{f_{2}}^{2}}{\mu
^{2}}]-\frac{3}{2}){\bigg )}, \notag
\end{align}%
and $\mu $ is the renormalization scale, being of the order of the
largest mass \cite{tamvakis}. Furthermore, we shall assume that
$m{L\simeq 1}$.\ This is required for the validity of perturbation
theory \cite{guth, s weinberg}.

It is well known that when one considers only twisted fermions and untwisted bosons in ${S}^{1}{\times R%
}^{3}$ (like in thermal field theories), vacuum contributions $\sim
$${L}^{-4}$ do not cancel and supersymmetry is spontaneously
broken.\ The non-cancelation occurs because bosons and fermions
satisfy different boundary conditions. In our model the field
content is such that cancelation of vacuum contributions is being
enforced, after having included all topologically inequivalent
allowed field configurations.

The leading order contribution to the one loop effective potential
is now given by \cite{dolan}:
\begin{align}
&V=m_{1}^{2}v^{2}+g_{1}^{2}m_{1}v^{3}+\frac{g_{1}^{2}v^{4}}{4} \label{pot}\\
&{-\frac{3(2m_{f_{1}}^{4}-m_{b_{1}}^{4}-m_{b_{2}}^{4})}{4096\pi ^{4}}-\frac{%
3(2m_{f_{2}}^{4}-m_{t_{1}}^{4}-m_{t_{2}}^{4})}{256\pi ^{4}}}\notag\\
&{+\frac{%
3(2m_{f_{1}}^{4}-m_{b_{1}}^{4}-m_{b_{2}}^{4}+2m_{f_{2}}^{4}-m_{t_{1}}^{4}-m_{t_{2}}^{4})%
}{128\pi ^{2}}}\notag \\
&{+\frac{(\gamma -\ln [4\pi ])(2m_{f_{1}}^{4}-m_{b_{1}}^{4}-m_{b_{2}}^{4})}{%
1024\pi ^{4}}+\frac{(\gamma +\ln [\frac{2}{\pi }%
])(2m_{f_{2}}^{4}-m_{t_{1}}^{4}-m_{t_{2}}^{4})}{64\pi ^{4}}}\notag \\
&{+\frac{(2m_{f_{1}}^{3}-m_{b_{1}}^{3}-m_{b_{2}}^{3})}{384L\pi ^{3}}-\frac{%
(2m_{f_{1}}^{2}-m_{b_{1}}^{2}-m_{b_{2}}^{2})}{768\pi ^{2}L^{2}}+\frac{%
(2m_{f_{2}}^{2}-m_{t_{1}}^{2}-m_{t_{2}}^{2})}{384\pi ^{2}L^{2}}}\notag\\
&{+\frac{2m_{f_{1}}^{4}\ln [Lm_{f_{1}}]-m_{b_{2}}^{4}\ln
[Lm_{b_{2}}]-m_{b_{1}}^{4}\ln [Lm_{b_{1}}]}{1024\pi ^{4}}}\notag \\
&{+\frac{2m_{f_{2}}^{4}\ln [Lm_{f_{2}}]-m_{t_{2}}^{4}\ln
[Lm_{t_{2}}]-m_{t_{1}}^{4}\ln [Lm_{t_{1}}]}{64\pi ^{4}}}\notag \\
&{-\frac{(2m_{f_{1}}^{4}\ln [\frac{m_{f_{1}}^{2}}{\mu ^{2}}]-m_{b_{2}}^{4}\ln [%
\frac{m_{b_{2}}^{2}}{\mu ^{2}}]-m_{b_{1}}^{4}\ln
[\frac{m_{b_{1}}^{2}}{\mu
^{2}}])}{64\pi ^{2}}}\notag\\
&{-\frac{(2m_{f_{2}}^{4}\ln [\frac{m_{f_{2}}^{2}}{\mu ^{2}}]-m_{t_{2}}^{4}\ln [%
\frac{m_{t_{2}}^{2}}{\mu ^{2}}]-m_{t_{1}}^{4}\ln
[\frac{m_{t_{1}}^{2}}{\mu ^{2}}])}{64\pi ^{2}}.\notag}
\end{align}%
Since relation (\ref{sp}) holds, the terms proportional to
$\frac{1}{L^{2}}$ cancel out \cite {love}.\ Also, the minimum of the
potential vanishes at ${v=}$0 and supersymmetry is
preserved.\ Indeed, upon expanding (\ref{pot}) for small values of ${v}$ we get:%
\begin{equation}
V\simeq m_{1}^{2}v^{2}+O(v^{3}).
\end{equation}%
In figure {1} we plot the effective potential for the limiting case
$mL=1$.\ The other
numerical values are chosen to be: ${m}_{1}{=}${200, }${m=}${7000, }${g}_{1}{%
=}${0.001, }${g=}${0.09, }${\mu =}${7000}.

Next, we introduce into the Lagrangian (\ref{test}) a term of the
form $g_{3}\chi ^{2}\varphi _{u_{2}}^{2}$, where ${g}_{3}$ a
dimensionless coupling constant ($\chi =v+\varphi
_{u_{1}}/\sqrt{2}$).\ This term, being non holomorphic and hard, breaks supersymmetry explicitly. Since $\chi $ develops a vev, $%
\varphi _{u_{2}}$ will acquire an additional mass term of the form $
g_{3}v^{2}$. This way, the masses of the fields now become:%
\begin{eqnarray}
m_{b_{1}}^{2} &=&m_{1}^{2}+3g_{1}m_{1}v+3g_{1}^{2}v^{2}/2 \\
\ m_{b_{2}}^{2} &=&m_{1}^{2}+g_{1}m_{1}v+g_{1}^{2}v^{2}/2+g_{3}v^{2}  \notag
\\
m_{t_{1}}^{2}\bigskip &=&m^{2}+4gmv+4g^{2}v^{2}+g^{2}m_{1}v/\sqrt{2}%
+g^{2}g_{1}v^{2}/4  \notag \\
\ m_{t_{2}}^{2} &=&m^{2}+4gmv-g^{2}m_{1}v/\sqrt{2}-g^{2}g_{1}v^{2}/4  \notag
\\
m_{f_{1}} &=&m_{1}+g_{1}v,\ m_{f_{2}}=m+2gv.  \notag
\end{eqnarray}%
As expected, supersymmetry is now broken and relation (\ref{str})
becomes,
\begin{equation}
2m_{f_{1}}^{2}-m_{b_{1}}^{2}-m_{b_{2}}^{2}=g_{3}v^{2},\
m_{t_{1}}^{2}+m_{t_{2}}^{2}=2m_{f_{2}}^{2}.
\end{equation}%
One can see that the supersymmetric minimum at $v=0$ is still
preserved. Indeed, $V$ can be
written as:%
\begin{equation}
V\simeq (m_{1}^{2}+\frac{g_{3}}{768\pi ^{2}L^{2}})v^{2}+O(v^{3}).
\end{equation}%
Upon closer examination, we can see that in the continuum limit, the
supersymmetric vacuum becomes metastable and a second non
supersymmetric vacuum appears. Including finite size corrections, we
see that for small ${L}$ the effective potential has a unique
supersymmetric minimum at $v=$0. As ${L}$ increases, a second
minimum develops, which becomes supersymmetric at the critical value
${L}_{c}{=}\frac{1}{21571}$.\ When ${L}\!>\!{L}_{c}$ the second
minimum breaks supersymmetry and becomes energetically more
preferable than the supersymmetric one \cite{linde,zeldovich}.\ This
said behavior of the potential is always valid whenever $g_{3}\gg g_{1}$ and for $%
\frac{m_{1}}{m}\ll g_{3}$. Using the same numerical values as
before, we plot the effective potential for ${g}_{3}{=}${0.5}, first
in the continuum limit (figure~2), and then including ${L}$
dependent corrections (figure~3).

Let us discuss the above results. $g_{3}$, $g_{1}$ are couplings
among the untwisted superfield, $g_{3}$ corresponding to the
supersymmetry breaking term. If the $g_{3}$ interaction is stronger
than $g_{1}$ and if the mass (${m}$) of the twisted superfield is
larger than the untwisted one (${m_{1}}$), then the following
picture occurs. For small length ${L}$ of the $S^{1}$ dimension
supersymmetry is not broken (figure~3). As ${L}$ grows larger, a
second minimum appears which is not supersymmetric
(${L}\!>\!{L}_{c}$). There exists a small barrier separating the two
minima (figure~3), and there is a possibility of barrier penetration
between them. This resembles the first order phase transition
picture.
\section*{\protect \Large Conclusions}
We have investigated the possibility of altering the spacetime
topology, without breaking ${N=1}$, ${d=4}$ supersymmetry.\ We have
explicitly demonstrated, by means of a toy model, that such a
construction is possible in the case of ${S}^{1}{\times R}^{3}$
topology. Introducing an explicit supersymmetry breaking term, we
find that, under suitable conditions, a second supersymmetric vacuum
appears. For small length of the $S^{1}$ dimension, only the
supersymmetric vacuum appears. As the length grows larger the second
minimum appears which, after a critical value, becomes non
supersymmetric. This picture resembles the first order phase
transition picture.

\section*{Acknowledgements}

V.O would like to thank I.Zois for useful conversations.

\noindent This research was co-funded by the European Union in the
framework of the Program PYTHAGORAS-I of the "Operational Program
for Education and Initial Vocational Training" (EPEAEK) of the 3rd
Community Support Framework of the Hellenic Ministry of Education,
funded by 25{\%} from national sources and by 75{\%} from the
European Social Fund (ESF).

\newpage
\begin{figure}[tbp]
\begin{center}
\epsfxsize=11cm\epsfysize=9cm\epsffile{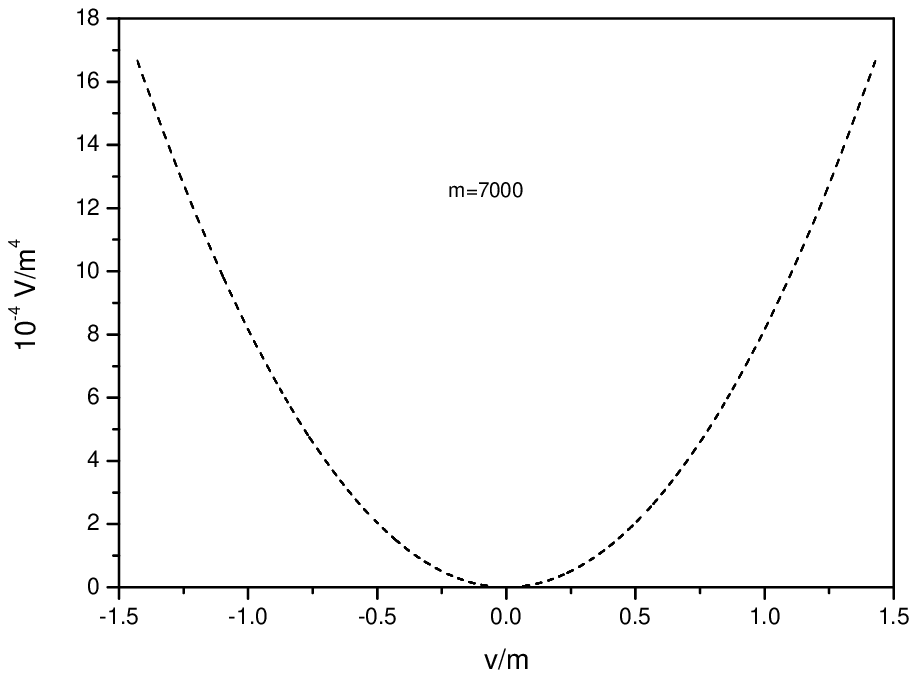}
\end{center}
\caption{The supersymmetric effective potential}
\end{figure}

\begin{figure}[tbp]
\begin{center}
\epsfxsize=11cm\epsfysize=9cm\epsffile{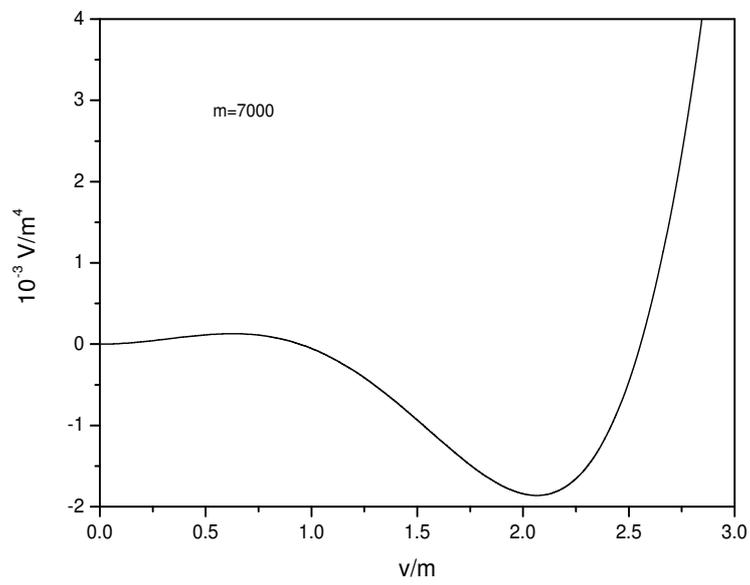}
\end{center}
\caption{The continuum effective potential}
\end{figure}

\begin{figure}[tbp]
\begin{center}
\epsfxsize=11cm\epsfysize=9cm\epsffile{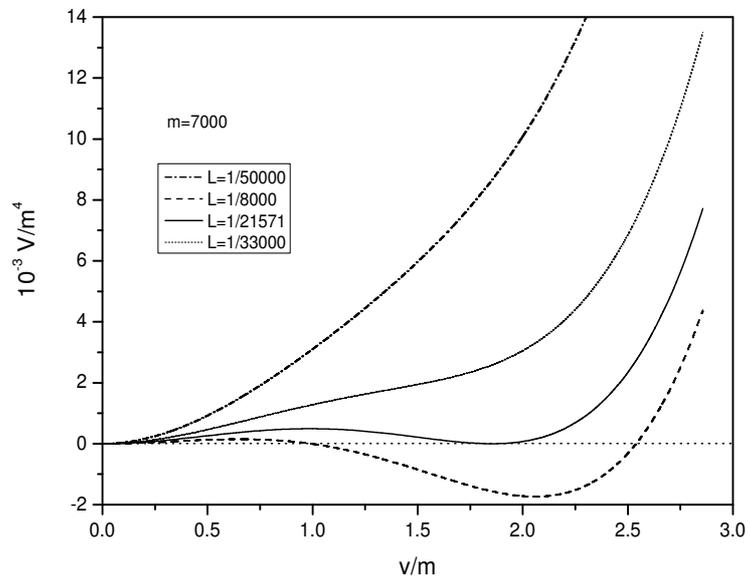}
\end{center}
\caption{The effective potential including finite size corrections}
\end{figure}

\end{document}